\documentclass[reprint, superscriptaddress, amsmath,amssymb, aps, prx]{revtex4-2}
\usepackage{graphicx}
\usepackage{dcolumn}
\usepackage{bm}
\usepackage{hyperref}
\usepackage{textgreek}

\usepackage{xcolor}
\usepackage{soul}

\begin{document}

\preprint{}

\title{Transport of topological defects in a biphasic mixture \\
of active and passive nematic fluids}

\author{Chaithanya K. V. S.}
\affiliation{School of Life Sciences, University of Dundee, Dundee DD1 5EH, United Kingdom}
\author{Aleksandra Arda{\v s}eva}
\affiliation{Niels Bohr Institute, University of Copenhagen, Blegdamsvej 17, Copenhagen, Denmark}
\author{Oliver J. Meacock}
\affiliation{Department of Fundamental Microbiology, University of Lausanne, Lausanne, Switzerland}
\author{William M. Durham}
\affiliation{Department of Physics and Astronomy, University of Sheffield, Sheffield, United Kingdom}

\author{Sumesh P. Thampi\thanks{Corresponding author}}
\email{sumesh@iitm.ac.in}
\affiliation{Department of Chemical Engineering, Indian Institute of Technology Madras, Chennai-36, India. }

\author{Amin Doostmohammadi\thanks{Corresponding author}}
\email{doostmohammadi@nbi.ku.dk}
\affiliation{Niels Bohr Institute, University of Copenhagen, Blegdamsvej 17, Copenhagen, Denmark}

\begin{abstract}
Collectively moving cellular systems often contain a proportion of dead cells or non-motile genotypes. When mixed, nematically aligning motile and non-motile agents are known to segregate spontaneously. However, the role that topological defects and active stresses play in shaping the distribution of the two phases remains unresolved. In this study, we investigate the behaviour of a two-dimensional binary mixture of active and passive nematic fluids to understand how topological defects are transported between the two phases and, ultimately, how this leads to the segregation of topological charges.  When the activity of the motile phase is large, and the tension at the interface of motile and non-motile phases is weak, we find that the active phase tends to accumulate $+1/2$ defects and expel $-1/2$ defects so that the motile phase develops a net positive charge. Conversely, when the activity of the motile phase is comparatively small and interfacial tension is strong, the opposite occurs so that the active phase develops a net negative charge. We then use these simulations to develop a physical intuition of the underlying processes that drive the charge segregation. Lastly, we quantify the sensitivity of this process on the other model parameters, by exploring the effect that anchoring strength, orientational elasticity, friction, and volume fraction of the motile phase have on topological charge segregation. As $+1/2$ and $-1/2$ defects have very different effects on interface morphology and fluid transport, this study offers new insights into the spontaneous pattern formation that occurs when motile and non-motile cells interact.
\end{abstract}

\maketitle
\section{Introduction}

Collective motility is observed in a wide diversity of systems where cells live in close proximity to each other, ranging from microbial communities in the environment to tumours within the human body. However, such collectively moving systems often contain a significant fraction of non-motile cells, commonly due to the loss of motility in some members or their inability to acquire it \cite{vlamakis2008control,gude2020bacterial,muok2021intermicrobial}. The interactions between motile and non-motile cells are crucial in shaping the dynamics, stability, and functionality of these systems, thus playing a key role in the prosperity of micro-ecosystems \cite{xu2019self,curatolo2020cooperative}. For instance, within biofilms, bacteria employ motility-based segregation as a strategy to evade exposure to antibiotics \cite{benisty2015antibiotic,zuo2020dynamic}. Similarly, cancer tumours present a varied composition, comprising both active and necrotic cells that vary in their motility \cite{aktipis2012dispersal,gallaher2019impact}. Understanding these interactions is crucial for unravelling the mechanisms driving tumour invasion and bacterial infection, including their treatment and resistance to therapies. Despite their prevalence, few studies have examined systems with both motile and non-motile phases, limiting our understanding. Here, we investigate the behaviour of a binary mixture of active and passive fluids using a biphasic active nematic framework to understand interactions between motile and non-motile phases.

\begin{figure}[htb!]
	\centering
	\includegraphics[width=0.95\linewidth]{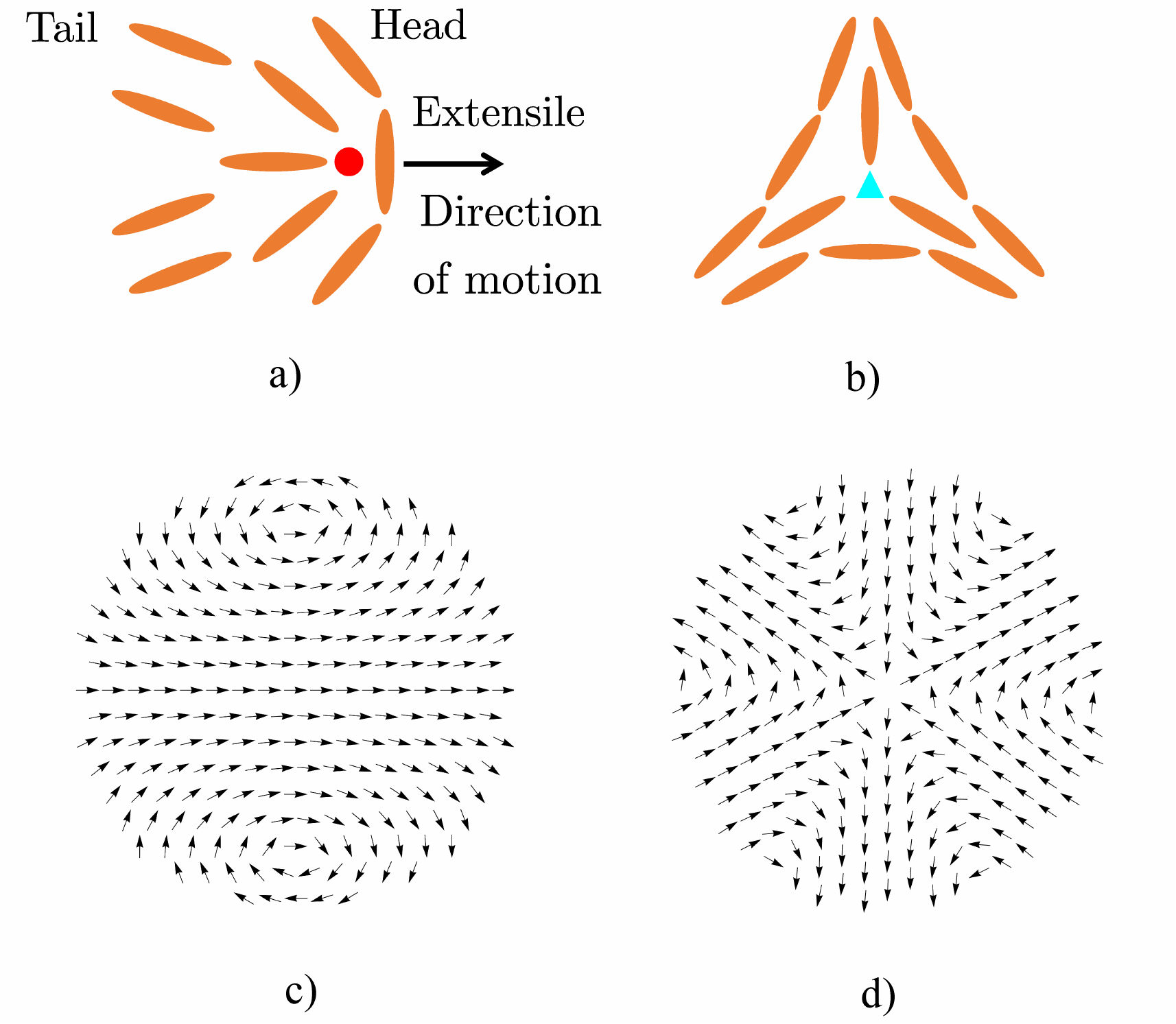}
	\caption{\textbf{Topological defects:} Schematic illustrating the director texture surrounding (a) $+1/2$ and (b) $-1/2$ topological defects observed in active nematic liquid crystals. The red circle denotes the core of the $+1/2$ defect, while the core of the $-1/2$ defect is represented by a cyan triangle.  Panels (c) and (d) illustrate the corresponding velocity field surrounding the $+1/2$ and $-1/2$ defects for extensile activity, respectively, computed using analytical expressions from ref.~\cite{giomi2014defect}.}
	\label{fig:defects}
\end{figure}

The constituents of many active systems, such as eukaryotic cells~\cite{Saw_2018,armengol2023epithelia}, bacterial colonies~\cite{Dell_2018}, microtubule--kinesin mixtures~\cite{Sanchez_2012}, and active filaments~\cite{Rui_2018} have nematic symmetry. Thus, active nematic theories have been instrumental in understanding
the large-scale flow patterns and coherent motion observed in these systems~\cite{Doostmohammadi_2018,partovifard2024controlling}. One of the remarkable features of two-dimensional active nematics is the formation of half-integer ($\pm1/2$) topological defects \cite{mur2022continuous}. These are regions of broken orientational order, as shown in Fig.~\ref{fig:defects}. 

Recently, there has been a growing interest in understanding the role of topological defects on the dynamics of a binary mixture of active nematic materials with varying levels of activity. For instance, \citet{Meacock2021} demonstrated how bacteria that move at different rates compete with one another when mixed together in colonies. Fast-moving cells can become trapped vertically in rosettes formed due to the merger of two $+1/2$ defects, which then allows slower-moving cells to outcompete them. Similarly, \citet{zhang2022topological} investigated defect-mediated morphogenesis of initially flat active-active interfaces using biphasic nematic theory and experiments based on Mardin–Darby canine kidney (MDCK) and mouse myoblast (C2C12) cells. They identified activity-mediated defect-interface interactions and studied the morphodynamics of an initially flat active-active interface, demonstrating the activity-dependent segregation of topological defects in a binary mixture of two active fluids. The passive liquid crystalline environment can also be pre-patterned to guide active nematic motion \cite{Genkin2017,Sciortino_2022}. Moreover, $+1/2$ defects have been shown to induce apoptotic cell extrusion in MDCK cell layers~\cite{Saw_2018}, and with multilayer formation in soil bacteria~\cite{copenhagen2020topological}, while $-1/2$ defects have been shown to contribute to hole formation in epithelial cell layers on soft substrates~\cite{sonam2023mechanical}.  A few studies probed the phase-separation of a binary mixture of active-isotropic fluids~\cite{bhattacharyya2023phase, caballero2022activity} and the dynamics of active–isotropic fluid interfaces~\cite{blow2014biphasic,adkins2022dynamics,coelho2019active,coelho2020propagation,coelho2023active}. These studies highlight the role of activity and nematic stresses on micro-phase separation, active anchoring at the fluid--fluid interface, and spatial distribution of topological defects. Furthermore, the origin of large-scale tissue flows during gastrulation in embryos~\cite{saadaoui2020tensile} and in cellular aggregates~\cite{yadav2022gradients,mehes2012collective,mccandlish2012spontaneous} is explained using the concepts of tissue interfacial tension, akin to the interfacial tension observed in fluid-fluid interfaces. Nevertheless, the intricate interplay between the forces of activity and interfacial tension in multi-species systems remains largely unexplored.

In this study, we employ the biphasic nematic framework to investigate the defect transport and the segregation of topological charges within a binary mixture of active-passive nematic fluids. We demonstrate that the charge of the active fluid can be tuned via the interplay between activity and interfacial tension. For high activities and low interfacial tension, we find the active fluid becomes positively charged due to the accumulation of active nematic within the cores of $+1/2$ defects and its depletion from $-1/2$ defect cores, a phenomenon that is consistent with experimental findings in monolayers composed of a single type of cells like epithelial cell layers~\cite{kawaguchi2017topological}, neural progenitor cells~\cite{copenhagen2020topological}, and soil bacteria~\cite{sonam2023mechanical}.
For larger interfacial tension and lower activity, the $+1/2$ defects are expelled from the active nematic, leading to a negatively charged active nematic and a positively charged passive nematic. Further, we establish the correlation between charge segregation and interface morphology. By systematically varying activity and interfacial tension strengths, we construct a phase diagram illustrating the net topological charge of the active nematic to elucidate the intricate interplay between the morphodynamics of interfaces and the transport of defects across these interfaces.  Moreover, we demonstrate distinct impacts of the anchoring strength, orientational elasticity, isotropic friction, and the initial concentration of the active fluid on charge segregation. The dependence of the charge segregation on these parameters is shown to align with their impact on the activity and number of topological defects.

This paper is organized as follows: In Sec.~\ref{sec:methods}, we provide details of the biphasic nematics modelling framework for a binary mixture of active and passive fluids. In Sec.~\ref{sec:results}, we present our findings on charge segregation. Specifically, in Sec.~\ref{sec:temporal_dyanmics}, we discuss the temporal dynamics of phase and charge segregation, while Sec.~\ref{sec:ac_st_effect}~\&~\ref{sec:mechanism} address the sensitivity of charge segregation to surface tension and activity, along with the governing mechanisms. Subsequently, in Sec.~\ref{sec:other_params}, we analyze the impact of interfacial anchoring, orientational elasticity, and isotropic friction on charge segregation. Finally, we explore the effect of the initial concentration of the active fluid on charge segregation in Sec.~\ref{sec:conc_effect}.

\noindent 

\section{Methods}\label{sec:methods}

We extend the two-dimensional lyotropic model, presented by \citet{blow2014biphasic}, to account for a biphasic fluid with interfacial tension. The two fluids are distinguished by a scalar order parameter, $\phi$, a measure of the relative concentration of each fluid. The nematic order in both fluids is described by a symmetric traceless tensor, $Q_{\alpha\beta} = S(2 n_\alpha n_\beta-\delta_{\alpha \beta})$, where $\mathbf{n}$ describes the director alignment and $S$ is the magnitude of the nematic order.

The energy density of the system is given by:
\begin{align}\label{eqn:free_energy}
    \mathcal{F} &=  \int \Bigg[\frac{A}{2}\phi^2(1-\phi)^2+\frac{1}{2}K\partial_\gamma \phi \partial_\gamma \phi+\\ \notag
    &\frac{1}{2}C(S^2-\frac{1}{2}Q_{\alpha\beta}Q_{\alpha\beta})^2+\frac{1}{2}L\partial_\gamma Q_{\alpha\beta} \partial_\gamma Q_{\alpha\beta}-\mu \phi\Bigg] d^2 \mathbf{r},
\end{align}
where $A$, $C$, $K$, and $L$ are positive constants.  The first term represents a double-well potential with minima at $\phi = 0$ (passive fluid) and $1$ (active fluid). The second term is the mixing term that penalizes the gradients in $\phi$. These two terms describe the phase separation of two fluids. The interfacial tension between the two fluids is given by $\gamma = \sqrt{AK}/6$~\cite{mueller2021phase}. The third term promotes nematic ordering, while the fourth term accounts for the nematic elasticity in the liquid crystalline energy of both fluids. Here $\mu$ is a Lagrange multiplier that conserves the integrated value of $\phi$, and $\mathbf{r}$ is the position vector. From here onward, we use Greek indices to represent Cartesian components, and repeated indices imply summation.

The order parameters, $\phi$ and $\mathbf{Q}$, evolve according to the advection--diffusion equations:
\begin{equation}\label{eqn:ADE_phi}
    \partial_t \phi + \nabla \cdot (\mathbf{u}\phi) = \Gamma_\phi \nabla^2\mu,
\end{equation}
\begin{equation}\label{eqn:ADE_Q}
    (\partial_t+\mathbf{u} \cdot \nabla)\mathbf{Q}-\mathbf{R} = \Gamma_Q \mathbf{H},
\end{equation}
where $\mathbf{u}$ is the velocity field, $\Gamma_\phi$ is the interface mobility parameter, which describes the rate at which $\phi$ responds to the gradients in the chemical potential, $\mu  = \frac{\partial \mathcal{F}}{\partial \phi}$. Similarly, $\Gamma_Q$ is the rotational diffusivity and $\mathbf{H}$ is the molecular field, defined as:
\begin{equation}\label{eqn:molecular_field}
    \mathbf{H} =  -\bigg(\frac{\partial \mathcal{F}}{\partial \mathbf{Q}}-(\mathbf{I}/2)\text{Tr}\frac{\partial \mathcal{F}}{\partial \mathbf{Q}}\bigg),
\end{equation}
where $\mathbf{I}$ is the identity matrix and \text{Tr} denotes the tensorial trace.
Unlike the order parameter, $\phi$, which only gets advected by the flow, the nematic constituents can rotate in response to the flow gradients. This is accounted by the co-rotation term:
\begin{align}\label{eqn:corotational_term}
R_{ij} = &(\xi D_{ik}+\Omega_{ik})\bigg(Q_{kj}+\frac{\delta_{kj}}{2}\bigg) \nonumber \\
&+\bigg(Q_{ik}+\frac{\delta_{ik}}{2}\bigg)(\xi D_{kj}-\Omega_{kj})  \\
&-2\xi\bigg(Q_{ij}+\frac{\delta_{ij}}{2}\bigg)Q_{kl}W_{lk} \nonumber,
\end{align}
where $D_{ij} = (\partial_j u_i+\partial_i u_j)/2$ and $\Omega_{ij} = (\partial_j u_i-\partial_i u_j)/2$ are the symmetric and anti-symmetric parts, respectively, of the velocity gradient tensor, $W_{ij} = \partial_i u_j$. The parameter $\xi$ quantifies the response of the director to the shear flow, and is related to the flow alignment parameter, $\lambda = \xi/(2S)$.

The fluid velocity evolves according to the Navier-Stokes equations:
\begin{align} \label{eqn:Navier_stokes}
   \nabla \cdot \mathbf{u} &= 0, \\
   \rho(\partial_t+ \mathbf{u} \cdot \nabla)\mathbf{u} &= \nabla \cdot \bm{\sigma}-\chi \mathbf{u}\nonumber,
\end{align}
where $\rho$ denotes the fluid density, $\bm{\sigma} = \bm{\sigma}_{\text{passive}} + \bm{\sigma}_{\text{active}}$ represents the stress tensor encompassing both active and passive contributions, and $\chi$ is the friction coefficient~\cite{thampi2014active}.

The passive contributions to stress in both fluids are given by, $\bm{\sigma}_{\text{passive}}=\bm{\sigma}_{\text{viscous}}+\bm{\sigma}_{\text{capillary}}+\bm{\sigma}_{\text{elastic}}$, 
\begin{equation}\label{eqn:Viscous_stress}
   \bm{\sigma}_{\text{viscous}} = 2\eta\mathbf{D},
\end{equation}
\begin{align} \label{eqn:Capillary_stress}
    \bm{\sigma}_{\text{capillary}}& = (\mathcal{F}-\mu \phi) \mathbf{I}-\nabla\phi\bigg(\frac{\partial \mathcal{F}}{\partial(\nabla \phi)}\bigg)\\
    &+\nabla\phi\bigg(\frac{\partial \mathcal{F}}{\partial(\nabla^2 \phi)}\bigg)-\nabla\nabla\phi\bigg(\frac{\partial \mathcal{F}}{\partial(\nabla^2 \phi)}\bigg) \nonumber
\end{align}

\begin{align} \label{eqn:elastic_stress}
  \bm{\sigma}_{\text{elastic}}& = -p \mathbf{I}-\xi\bigg(\mathbf{H}(\mathbf{Q}+\mathbf{I}/2)+(\mathbf{Q}+\mathbf{I}/2)\mathbf{H}\\ \notag
    &-2(\mathbf{Q}+\mathbf{I}/2)\text{Tr}(\mathbf{Q}\mathbf{H})\bigg)+\mathbf{Q}\mathbf{H}\\ \notag
    &-\mathbf{H}\mathbf{Q} - \nabla\mathbf{Q}\bigg(\frac{\partial\mathcal{F}}{\partial(\nabla \mathbf{Q})}\bigg),
\end{align} 
where $\rho$ is the fluid density, $\eta$ is the fluid viscosity, $p$ is the bulk pressure.

The active contribution to the stress in the active fluid ($\phi=1.0)$ is given by,
\begin{equation} \label{eqn:active_stress}
    \bm{\sigma}_{\text{active}} = -\zeta \mathbf{Q},
\end{equation}
where, $\zeta$ represents the activity coefficient. A positive $\zeta$ corresponds to an extensile material, while a negative $\zeta$ corresponds to a contractile material.

The coupled equations for fluid velocity (eqn.~\ref{eqn:Navier_stokes}), the nematic (eqn.~\ref{eqn:ADE_Q}) and phase--field order parameters (eqn.~\ref{eqn:ADE_phi}) are solved using hybrid lattice--Boltzmann method~\cite{marenduzzo2007steady,carenza2019lattice} with periodic boundary conditions on square domain $L \times L$ of size $L = 500$.  The frictional force $-\chi \mathbf{u}$ is incorporated as a force density in the lattice Boltzmann scheme.  Simulations are performed on a two-dimensional Cartesian mesh of size $500 \times 500$ with periodic boundary conditions. The spatial and temporal resolutions are chosen as unity. The parameters used in simulations are listed in Table \ref{t:contParams}. Simulations are initialized with a quiescent velocity field ($\mathbf{u} = 0$), and the phase--field value, $\phi (t = 0) = 0.50$, indicating a mixture with equal volume fraction of active fluid, $\phi_a = \phi(t=0)$, and passive fluid, $\phi_p = 1-\phi(t=0)$. Additionally, the nematic director field is initialized close to the uniformly oriented state, $\mathbf{n} = \mathbf{e}_x$, using a random seed.

\begin{table}[bth]
\caption{Values of the parameters used in the biphasic model}
\label{t:contParams}
\centering
\begin{tabular}{ll} 
 \hline\noalign{\smallskip}
Parameter & Numerical value \\
 \noalign{\smallskip}\hline\noalign{\smallskip}
 Kinematic viscosity, $\mu = \eta/\rho$ & $0.167$ \\
 Rotational diffusivity, $\Gamma_Q$ & $0.1$ \\
 Mobility, $\Gamma_{\phi}$ & $0.1$ \\ 
 Tumbling parameter, $\lambda$ & $0.3$ \\
 Surface tension, $\gamma$ & $0.0186-0.0559$ \\
 Parameter in free energy (Eqn.~\ref{eqn:free_energy}), $C$ & $0.5$\\
 orientational elasticity, $L$ & $0.00-0.25$ \\
 Activity, $\zeta$ & $0.00-0.30$\\
 Friction coefficient, $\chi_0$ & $0.0-10$ \\
 \hline\noalign{\smallskip}
\end{tabular}
\end{table}

To vary the surface tension $\gamma = \sqrt{AK}/6$, the parameter $A$ is fixed at $0.5$, while $K$ is varied over the range $0.025-0.2250$. The results shown represent an average over three distinct simulations, each initialized with a different random seed that prescribes the random distribution of the director field, $\mathbf{n}$.

\section{Results and discussion}

\label{sec:results}

\begin{figure*}[!ht]
\centering
	\includegraphics[width=0.95\linewidth]{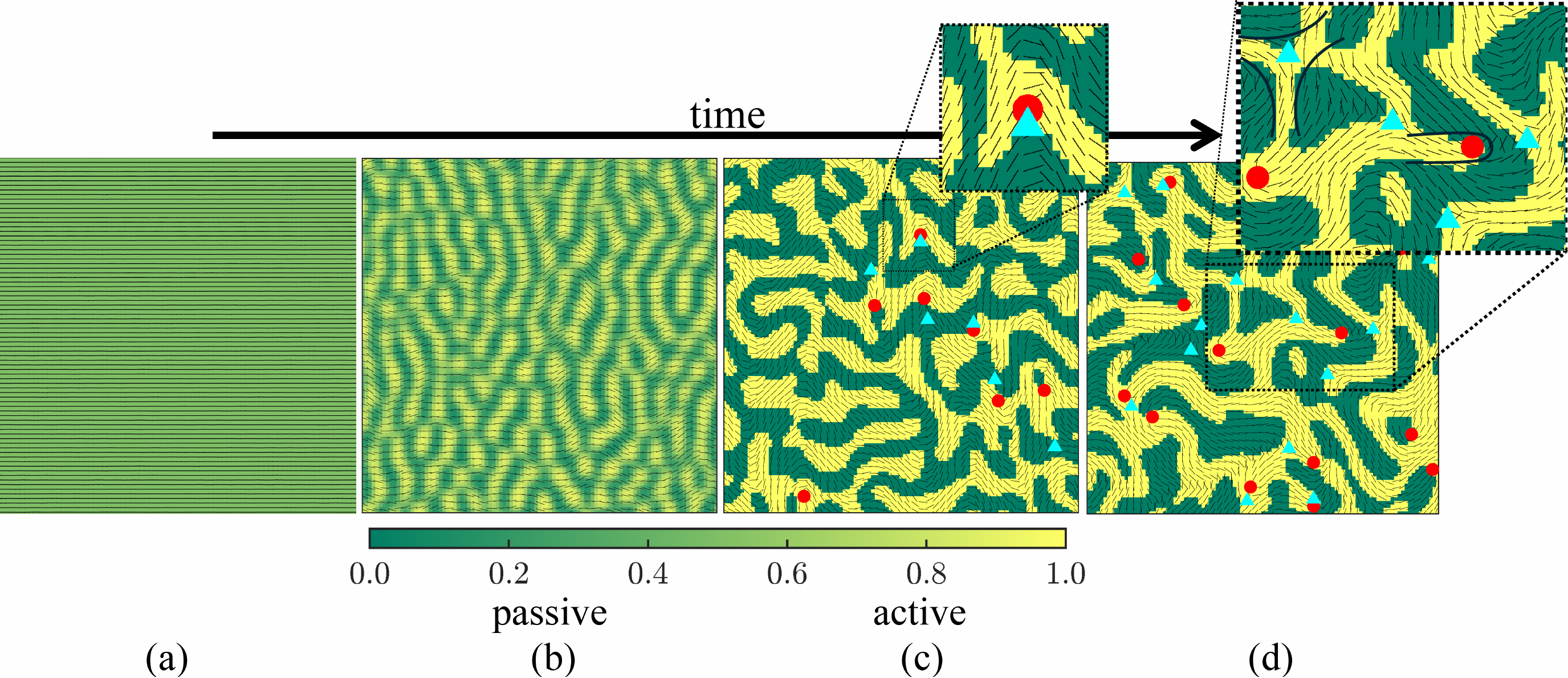}
	\caption{\textbf{Phase segregation:} The time evolution of the phase field variable ($\phi$) for $\zeta = 0.15$ and $K=0.05$, with overlaid director field (black lines) and topological defects ($+1/2$ depicted as red circles, $-1/2$ as cyan triangles). The snapshots correspond to time instances: (a) $t = 0$, (b) $t = 900$, (c) $t=1800$, and (d) $t = 25000$ simulation steps. An area of $100 \times 100$ within a larger domain sized $500 \times 500$ is shown for clarity. The zoomed-in region in (c) illustrates typical defect pair generation, and (d) shows the fluid-fluid interface morphology around the defects. }
	\label{fig:time_evol}
\end{figure*} 

\subsection{Temporal dynamics}

\label{sec:temporal_dyanmics}
\subsubsection{Phase segregation}

\label{sec:phase_seg}

We start by looking at the temporal evolution of the active-passive phases, as depicted in Fig.~\ref{fig:time_evol}. Initially, the system is in a fully mixed state with $\phi = 0.50$, and the director is nematically aligned, albeit with slight fluctuations (Fig.~\ref{fig:time_evol} (a)). The double-well potential of the phase field variable $\phi$ drives the phase separation process, promoting the formation of distinct active and passive phases characterized by $\phi = 1.0$ and $\phi = 0.0$, respectively (Fig.~\ref{fig:time_evol}(b)). Simultaneously, the interfacial tension acts to minimize the length of the interface between these phases, facilitating the demixing process. In contrast to mixtures of passive isotropic fluids which segregate and stabilize into a minimum energy equilibrium configuration \cite{cahn1965phase,datt2015morphological}, in the active nematic system, domains of active and passive fluids display a dynamic behaviour; the domains continuously break up and reform. Additionally, the presence of extensile active flows cause these domains to elongate parallel to the director field~\cite{blow2014biphasic,bhattacharyya2023phase} (Fig.~\ref{fig:time_evol} (d)). While interfacial tension drives the demixing of the active and passive fluids by minimizing the length of the interface between the two fluids, activity drives fluid mixing by elongating the interface between them. This interplay between interfacial tension and activity significantly influences the dynamics and morphology of the phase-separated domains.

Furthermore, the active flows induce local distortions in the nematic order, thereby promoting the formation of half--integer topological defects~\cite{blow2014biphasic}. As time progresses, we observe the spontaneous emergence of these defects (Fig.~\ref{fig:time_evol}(c)). Subsequently, these defects traverse across the fluid-fluid interface, enabling a continuous exchange of topological defects between the passive and active fluids  (Fig.~\ref{fig:time_evol} (d)). This behaviour contrasts with that of a single nematic fluid or a nematic-isotropic fluid mixture, where defects typically remain localized within the active fluid \cite{blow2014biphasic}.

Additionally, Fig.~\ref{fig:time_evol} (d) illustrates the morphology of the fluid--fluid interface near topological defects. The director field associated with a $+1/2$ defect promotes the formation of a comet-shaped interface, while a $-1/2$ defect induces a trefoil-shaped interface. This occurs because the extensile active flows elongate the fluid domains parallel to the director field. Moreover,  Fig.~\ref{fig:time_evol} (d) highlights the occurrence of effective anchoring, where the director field aligns parallel to the fluid--fluid interface, as observed in a nematic--isotropic mixture~\cite{blow2014biphasic}.  

\subsubsection{Charge segregation}

 \begin{figure} [h!]
	\centering
	\includegraphics[width=0.85\linewidth]{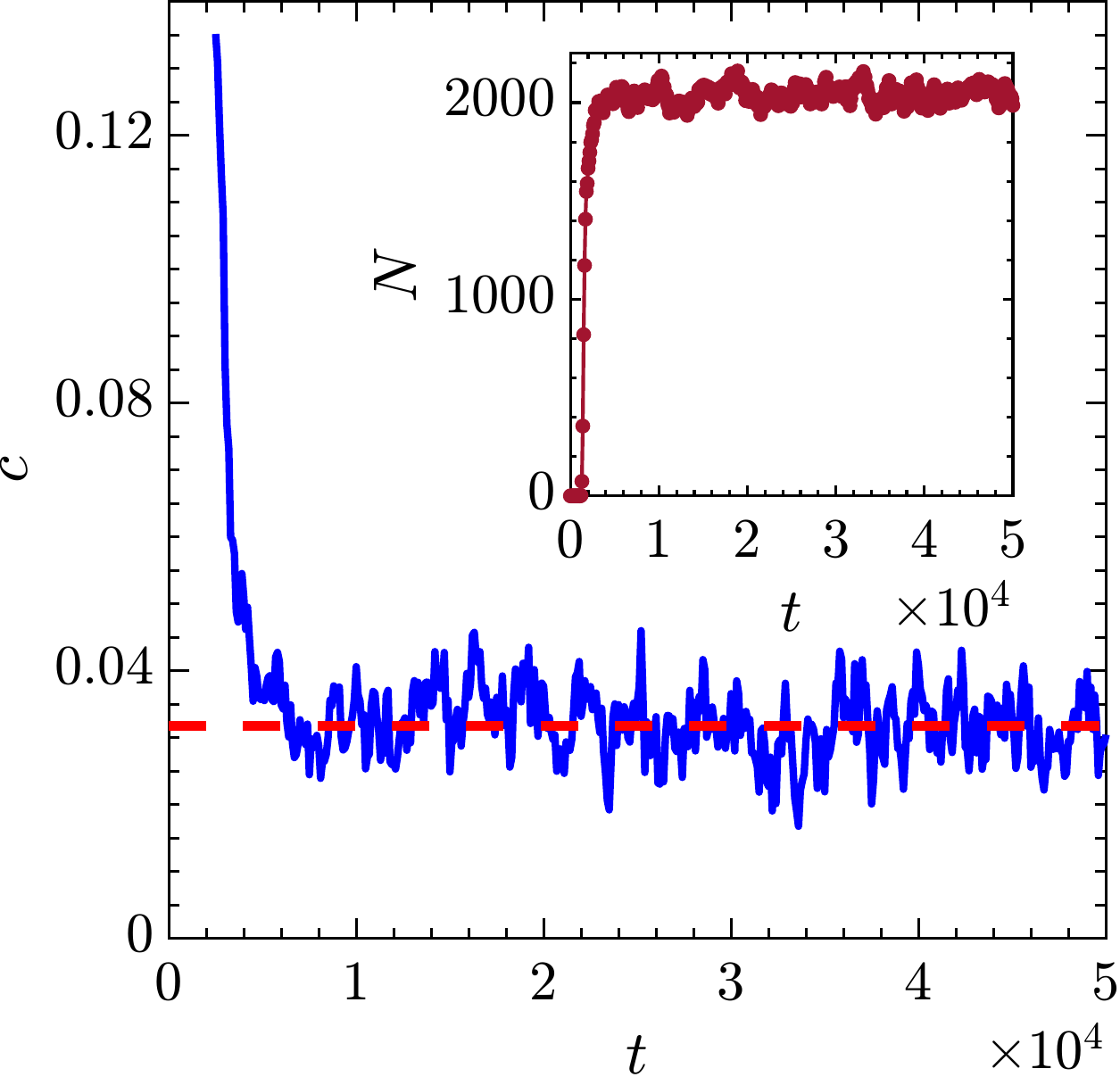}
	\caption{\textbf{Charge segregation:} The time evolution of the charge of the active fluid for $\zeta = 0.15$ and $K = 0.05$. The red dashed line indicates the average charge, corresponding to a dynamical steady state. Inset shows the temporal evolution of the total number of defects  ($N$) in the system.}
	\label{fig:cvst}
\end{figure}

In active nematic fluids, half-integer defects ($\pm 1/2$) are always generated in pairs, maintaining the overall charge neutrality of the system \cite{Doostmohammadi_2018}. In contrast, in the present case, the defects can traverse between the active and passive fluids; their exchange can be uneven, leading to disruptions in the charge neutrality within each phase. Consequently, one type of defect (either $+1/2$ or $-1/2$) may predominate in a given fluid, resulting in a net positive or negative charge. This phenomenon is referred to as the \textit{charge segregation} \cite{zhang2022topological}.

We quantify the charge segregation in terms of the average charge of the active fluid defined as,
\begin{equation}\label{eqn:charge_def}
c = \frac{\phi_a q_{\text{a}}}{N}.
\end{equation}

Here, $q_{\text{a}}$ is the net charge of the active fluid, defined as the sum of positive and negative topological charges of defects present in the active fluid, and $N = \phi_a N_a+\phi_p N_p$ denotes the total number of defects in the system, encompassing both the number of defects in active ($N_a$) and passive ($N_p$) fluids. When the number of $+1/2$ defects predominate over $-1/2$ defects in the active fluid, $c$ assumes a positive value ($c>0$), signifying a positively charged active fluid. Conversely, if $-1/2$ defects prevail in the active fluid compared to $+1/2$ defects, $c$ takes a negative value ($c<0$), indicating a negatively charged active fluid. Furthermore, the charge of the passive fluid is complementary; thus, $c>0$ denotes a negatively charged passive fluid, whereas $c<0$ implies a positively charged passive fluid.

Figure~\ref{fig:cvst} illustrates the temporal evolution of (i) the total number of defects in the system (inset of Fig.~\ref{fig:cvst}) and (ii) the charge of the active fluid ($c$) for $\zeta = 0.15$ and $K = 0.05$. Initially, the total number of defects increases rapidly, eventually stabilizing into a dynamic steady state, with number of defects fluctuating about a mean value. On the other hand, the charge of the active fluid starts at a high value and gradually decreases until it reaches a dynamic steady state. The charge of the active fluid fluctuates around this mean positive value indicated by the red dashed line in Fig.~\ref{fig:cvst}. The predominance of positive charge suggests the presence of larger number of $+1/2$ defects compared to $-1/2$ defects in the active fluid. This phenomenon occurs because defect pairs are usually generated in such a way that the $+1/2$ defect forms within the bulk of the active fluid and is oriented away from the interface, while the $-1/2$ defect forms relatively near the interface. This distribution arises because regions of large curvature are favourable for the formation of $+1/2$ defects, whereas regions of small curvature are more suitable for $-1/2$ defects~\cite{metselaar2019topology}.
As defects are continuously exchanged between the active and passive fluids, the charge of the active fluid stabilizes, fluctuating around a mean value. The fluctuations in the charge suggest that the system reaches a dynamic steady state, marked by the continuous exchange of topological defects between the two fluids. Similarly, the fluctuations in the total number of defects result from the continuous process of defect annihilation and creation. This behaviour contrasts with that observed in prepatterned stationary activity gradients with zero curvature~\cite{shankar2019hydrodynamics,ronning2023defect}. In such scenarios, $+1/2$ defects typically exhibit orientational polarization and tend to accumulate on the passive side of the interface due to weak mobility, while $-1/2$ defects remain within the active fluid. However, in this study, the activity is not prepatterned; instead, the spatiotemporal dynamics of activity are coupled to the evolution of the phase-field parameter, $\phi$.

\subsection{Sensitivity of charge segregation to interfacial tension and activity}
\label{sec:ac_st_effect}

 \begin{figure} [!htb]
	\centering\includegraphics[width=0.85\linewidth]{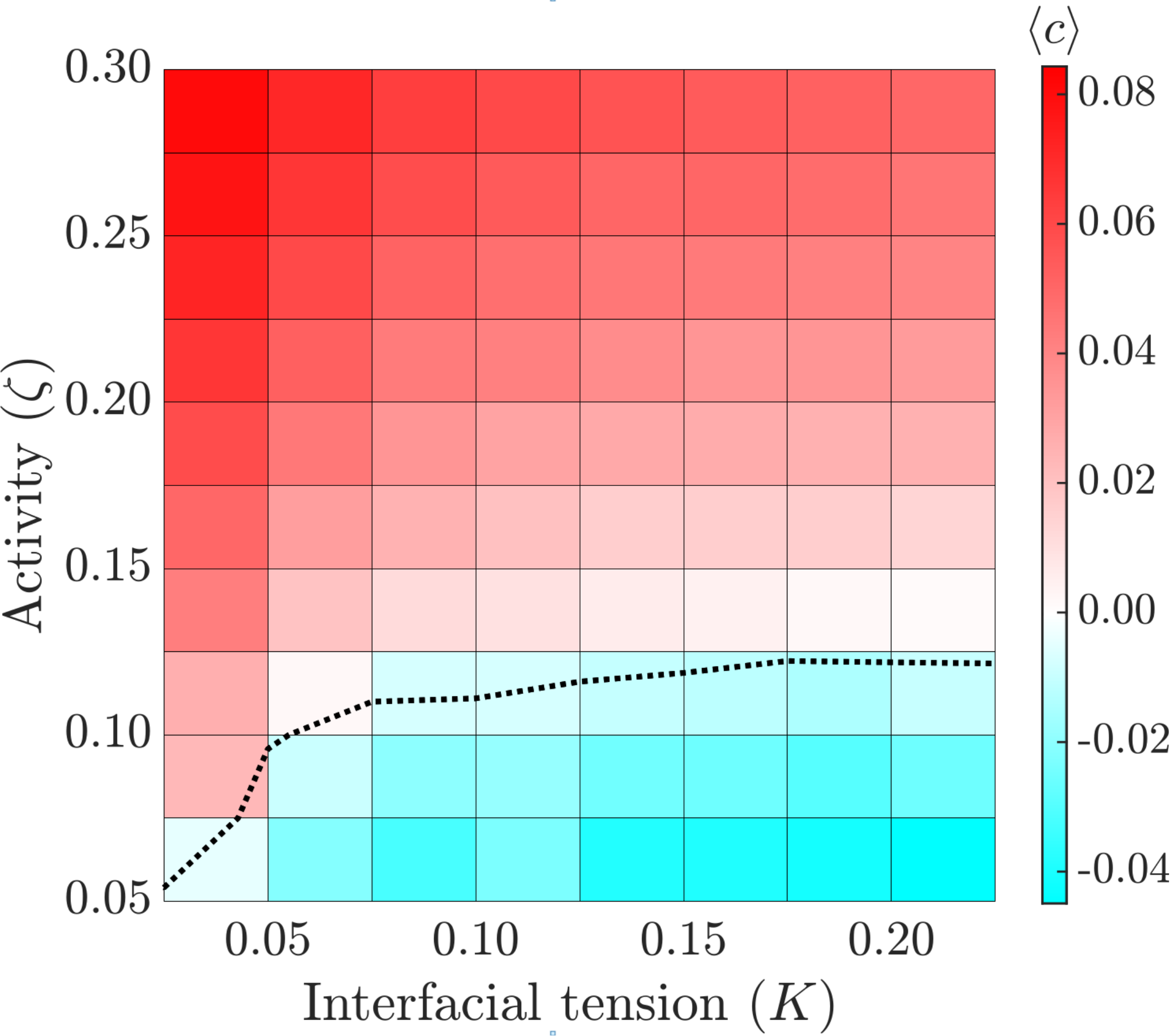}
	\caption{\textbf{Sensitivity of active fluids' charge to activity and interfacial tension
 :} Phase diagram depicting the dependence of average charge of active fluid on activity ($\zeta$) and interfacial tension ($K$).The black dotted line marks the region where the average charge ($<c>$) equals zero, indicating a change in the sign of the active fluids' charge.}
	\label{fig:state_diagram}
\end{figure}

To assess the dependence of charge segregation on activity and interfacial tension, we define the time-averaged charge of the active fluid as follows:
\begin{equation}
    \langle c \rangle = \frac{1}{T} \sum_{t=t_i}^{t_f} c(t).
\end{equation}
Here, we calculate the average charge between $t_i = 3.5 \times 10^{4}$ and $t_f = 5 \times 10^{4}$  in steps of $\Delta t = 100$, i.e., $T = 150$.
Figure~\ref{fig:state_diagram} depicts the dependence of the average charge of the active fluid, $\langle c \rangle$, on interfacial tension characterized in terms of $K$ and activity, $\zeta$.  When interfacial tension is maintained constant, an increase in activity leads to an increase in the charge of the active fluid, eventually transitioning from negative to positive. Conversely, with constant activity, the charge of the active fluid decreases as interfacial tension increases. The plot illustrates that it is not solely due to the activity but rather the interplay between activity and interfacial tension that determines the charge of the active fluid. Specifically, low activity combined with high interfacial tension yields a negatively charged active fluid, while high activity combined with low interfacial tension results in a positive charge. Additionally, the critical activity at which the charge changes sign from negative to positive depends on the interfacial tension, as indicated by the black dotted curve in Fig.~\ref{fig:state_diagram}. The critical activity for the charge reversal increases with increase in interfacial tension.

\subsection{Defect migration across phases leads to charge segregation}
 \label{sec:mechanism}
To unveil the mechanisms underlying charge segregation, we explore various pathways of defect transport between the two fluids. Specifically, we look at the interactions between the $\pm 1/2$ defects and the fluid-fluid interface.

\subsubsection{Transport of $+1/2$ defects}

 \begin{figure*}[!htb]
	\centering
	  \includegraphics[width =0.95\linewidth ]{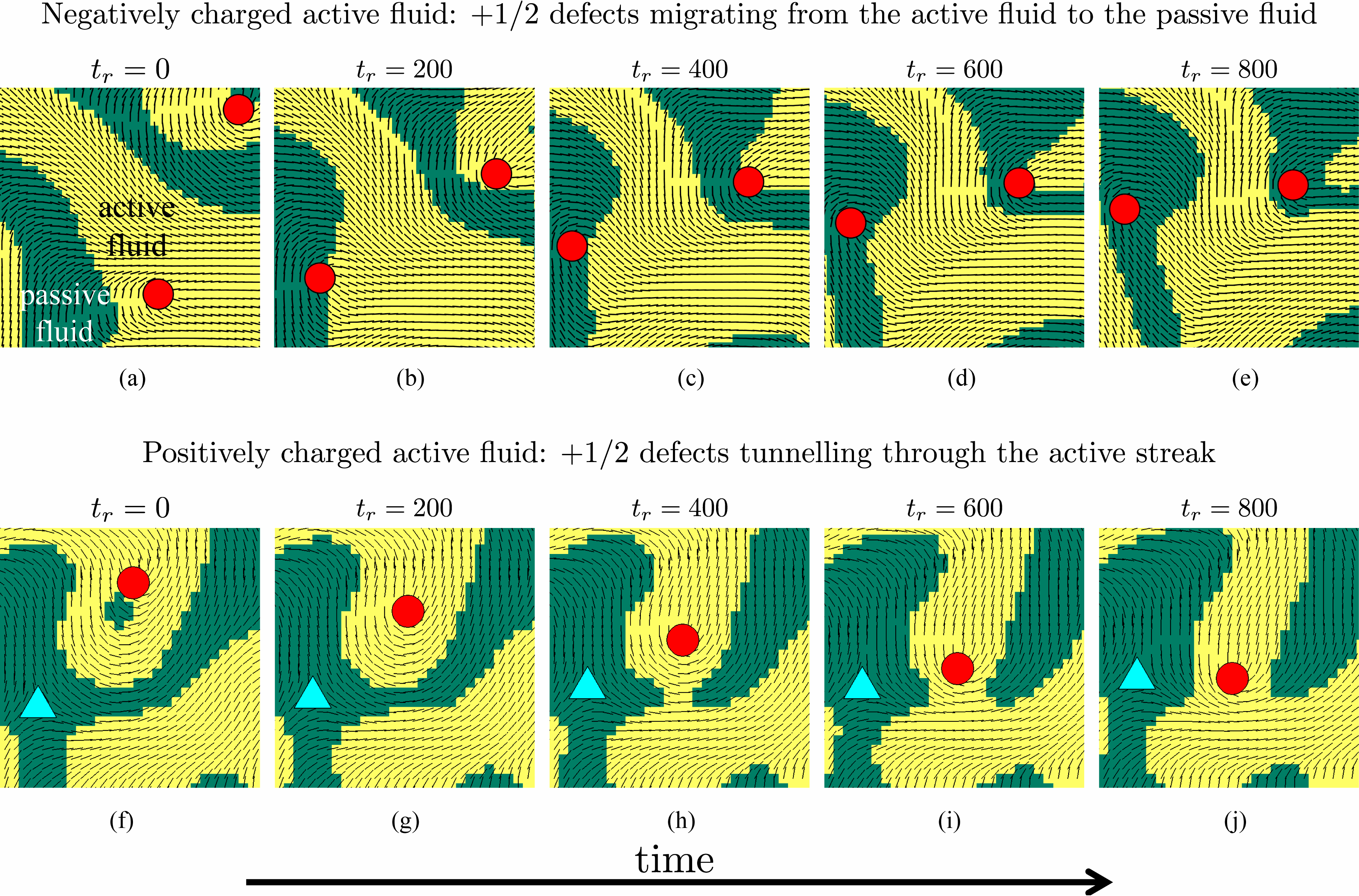}
	\caption{\textbf{Transport of $+1/2$ defects in a binary mixture of active-passive nematic fluids:} The $+1/2$ defect drives the formation of active fluid streaks, countered by interfacial tension. First row (a)-(e) corresponds to $\zeta = 0.100$ and $K = 0.225$: When interfacial tension dominates over activity, $+1/2$ defects exit the active fluid and migrate into the passive fluid. Second row (e)-(f) corresponds to $\zeta =0.20 $ and $K = 0.10 $: When activity dominates over interfacial tension, $+1/2$ defects tunnel through the streaks, carrying the streak along, and merge with another active streak. $t_r$ represents the rescaled simulation time, with its value set to $0$ for (a) and (f).}
	\label{fig:pd_mech}
\end{figure*}

First, we focus on the behaviour of $+1/2$ defects near a fluid--fluid interface, depicted in the Fig.~\ref{fig:pd_mech}. The first row (Fig.~\ref{fig:pd_mech}(a)-(e)) shows a representative scenario, which is prevalent when interfacial tension is high and activity is low. The $+1/2$ defects are generated in the bulk of the active fluid (Fig.~\ref{fig:pd_mech}(a)) and migrate toward the fluid-fluid interface due to inherent motility (Fig.~\ref{fig:pd_mech}(a) \& (c)). The director field around a $+1/2$ defect locally deforms the interface to a comet-like shape (Fig.~\ref{fig:time_evol}(d)). Simultaneously, the associated source-sink velocity field (Fig.~\ref{fig:defects} (c)) drives the interface in the same direction as the defect, forming an active streak. Interfacial tension, in turn, resists the deformation of the interface. Therefore, when interfacial tension dominates the activity, the elongation of the streak is limited, and $+1/2$ defects migrate towards the tip of the active streak and eventually exit the active fluid and enter the passive fluid, as shown in Fig.~\ref{fig:pd_mech} (d) \& (e). 

On the other hand, the second row of Fig.~\ref{fig:pd_mech} illustrates the scenario in which activity dominates over interfacial tension. Under these conditions, there is minimal resistance to streak deformation and mobility. Consequently, the $+1/2$ defects tunnel through the active streak (Fig.~\ref{fig:pd_mech} (f) \& (g)), effectively advecting the interface and merging with another active streak (Fig.~\ref{fig:pd_mech} (h)). Subsequently, the $+1/2$ defect and the corresponding active streak integrate into another active streak (Fig.\ref{fig:pd_mech} i)--(j)). Through this pathway, the $+1/2$ defects predominantly traverse within the active fluid through the continuous remodelling of fluid phases.

\subsubsection{Transport of $-1/2$ defects}

Here, we examine the behaviour of $-1/2$ defects near a fluid-fluid interface. Unlike $+1/2$ defects, $-1/2$ defects are non-motile and are advected by the fluid as passive tracers. Therefore, their impact on interface morphology and charge segregation is notable only when they are located close to the fluid-fluid interface.

 \begin{figure*} [!htb]
	\centering
	\includegraphics[width =0.95\linewidth ]{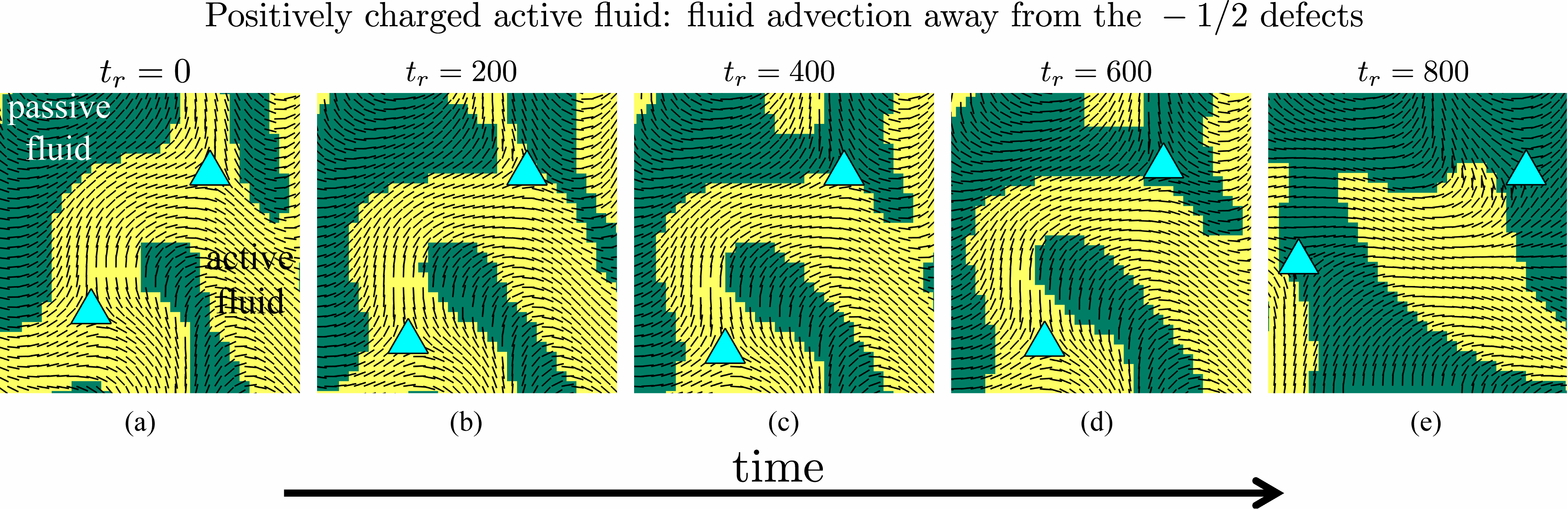}
	\caption{\textbf{Transport of $-1/2$ defects in a binary mixture of active-passive nematic fluids:} The active fluid assumes a trefoil shape near the $-1/2$ defects, and eventually, the fluid is advected away from the defect core. $t_r$ represents the rescaled simulation time, with its value set to $0$ for (a), and $\zeta = 0.20$, $K = 0.10$.}
	\label{fig:md_behaviour}
\end{figure*}

When interfacial tension is high and activity is low, the large size of the fluid domains, combined with the weak flow, causes the non-motile $-1/2$ defects to primarily reside within the bulk of the active fluid. Conversely, when activity dominates over interfacial tension, the $-1/2$ defects tend to interact more prominently with the fluid--fluid interface. Figure~\ref{fig:md_behaviour} depicts a scenario when $-1/2$ defect is located close to the fluid--fluid interface. The director field around the $-1/2$ defects and associated hexapolar velocity field (Fig.~\ref{fig:defects}~(b)\&(d)) locally deforms the interface to a trefoil--like shape (Fig.~\ref{fig:md_behaviour} (a) \& (b)) as described in Sec.~\ref{sec:phase_seg}. However, the active fluid eventually drifts away from the defect core along the trefoil axes, driven by extensile activity as depicted in Fig.~\ref{fig:md_behaviour}~(c)-(d), ultimately entering the passive fluid. The efficacy of the fluid depletion from the $-1/2$ defects increases with activity and decreases with interfacial tension.

In summary, increasing activity and decreasing interfacial tension lead to a positively charged active fluid. In these conditions, the $+1/2$ defects traverse by joining the active streaks without entering the passive fluid. Meanwhile, the active fluid gradually depletes from $-1/2$ defects, causing them to enter the passive fluid. Conversely, decreasing activity and increasing interfacial tension yield a negatively charged active fluid. Here, $+1/2$ defects exit the active fluid due to strong resistance to streak elongation, while $-1/2$ defects predominantly remain within the active fluid's bulk.

\subsection{Sensitivity to model parameters}
\label{sec:other_params}

\subsubsection{Effect of anchoring strength}

\begin{figure} [!htb]
    \centering
    \includegraphics[width=0.90\linewidth]{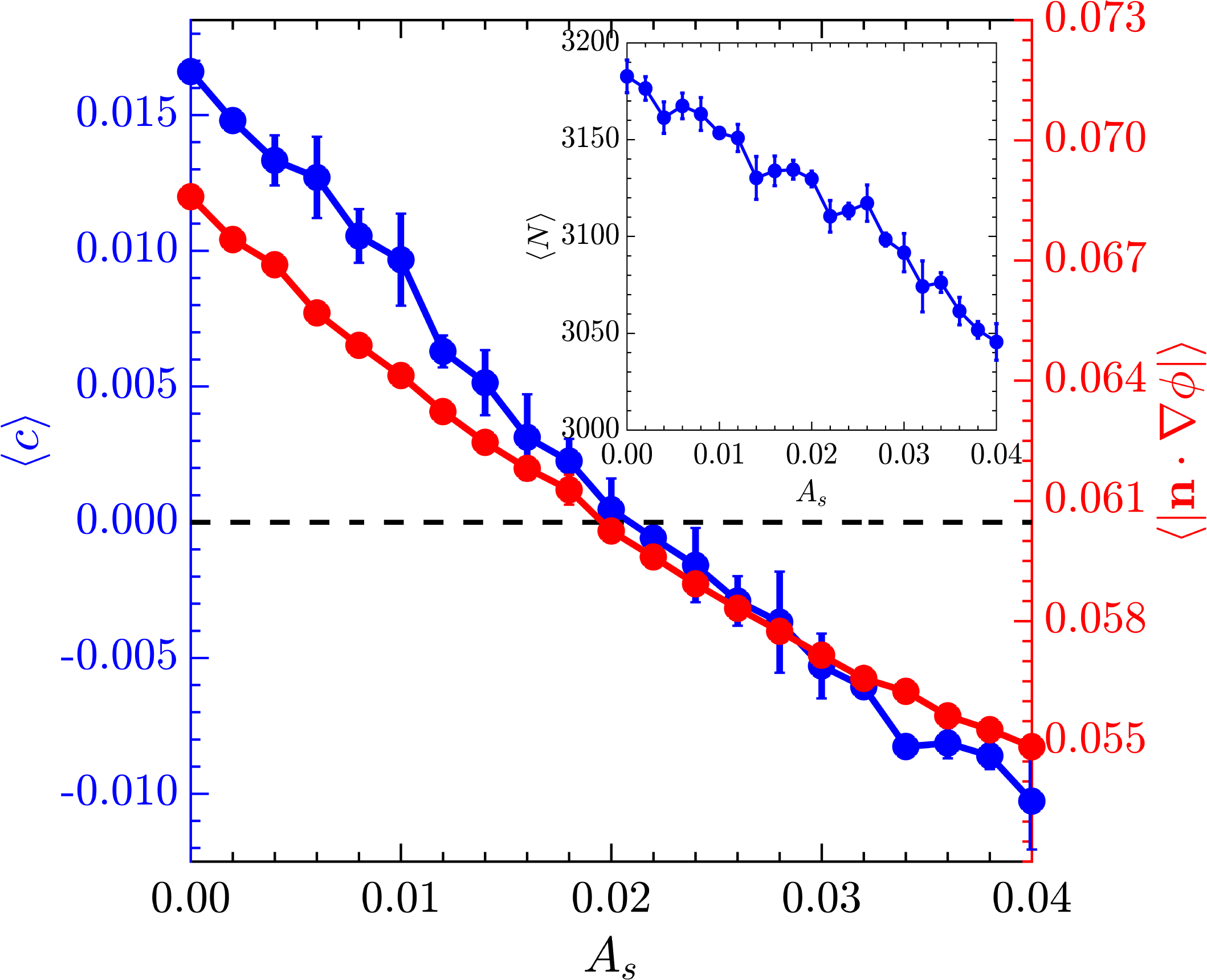}
     \caption{\textbf{Effect of anchoring strength, $A_S$:} Variation in the charge of the active fluid ($\langle c \rangle$) and the active anchoring of the director field at the interface ($\langle |\mathbf{n}\cdot \nabla\phi|\rangle$) to anchoring strength for $\zeta = 0.15$ and $K = 0.15$. Inset shows the variation in the total number of defects ($\langle N \rangle$) in the system with $A_s$. Error bars represent the standard deviation.}
    \label{fig:anchoring_effect_c_ngphi}
\end{figure} 
Next, we examine the impact of interfacial anchoring on charge segregation. Interfacial anchoring has been observed in the bacterial colonies \cite{doostmohammadi2016defect} and interacting multicellular monolayers \cite{zhang2022topological}. This anchoring occurs due to the rotation of the nematic director in response to the active force tangential to the interface \cite{blow2014biphasic,yadav2022gradients}. Consistent with these findings, we observe active anchoring in the biphasic system, where the director field tends to align parallel to the interface, as described in Sec.~\ref{sec:phase_seg}. However, the specific orientation of the director field at the interface can be enforced via the free energy term $F_{a}$ \cite{zhang2022topological},
\begin{equation}
    F_{a} = \int_{A} A_S \partial_\alpha \phi \partial_\beta \phi Q_{\alpha\beta} dA,
\end{equation}
where $A_S$ represents the anchoring strength, and $A_S > 0$ enforces tangential anchoring of the director field at the fluid-fluid interface.
Figure~\ref{fig:anchoring_effect_c_ngphi} illustrates that the charge of the active fluid decreases with increasing anchoring strength (blue curve). To understand the dependence of charge on anchoring strength, we measure the extent of alignment of the director field with the interface in terms of $\mathbf{n}\cdot\nabla\phi$. When the director field is aligned parallel to the interface, $\mathbf{n}\cdot\nabla\phi = 0$, indicating perfect alignment. Any deviation from zero suggests that the director field is oriented at an angle with the interface, indicating imperfect alignment. As expected, the average value of $|\mathbf{n}\cdot\nabla\phi|$ decreases with an increase in anchoring strength, indicating a greater alignment of the director field with the interface (Figure~\ref{fig:anchoring_effect_c_ngphi} - red curve).

This enhanced alignment, coupled with the elongation of interfaces due to extensile activity, results in a decrease in the number of topological defects, as depicted in the inset of Fig.~\ref{fig:anchoring_effect_c_ngphi}. Therefore, the effect of activity weakens with an increase in anchoring strength, resulting in a decrease in the charge of the active fluid. Additionally, an increase in anchoring strength increases the resistance to the transport of topological defects across the interface. Given that the passive fluid is predominantly governed by nematic elasticity, the resistance to the $+1/2$ defects migrating from the passive to active fluid surpasses that in the opposite direction. Consequently, the $+1/2$ defects tend to remain within the passive fluid, leading to a negatively charged active fluid at high anchoring strength.

 \subsubsection{Effect of orientational elasticity}

\begin{figure} [!htb]
    \centering
    \includegraphics[width=0.80\linewidth]{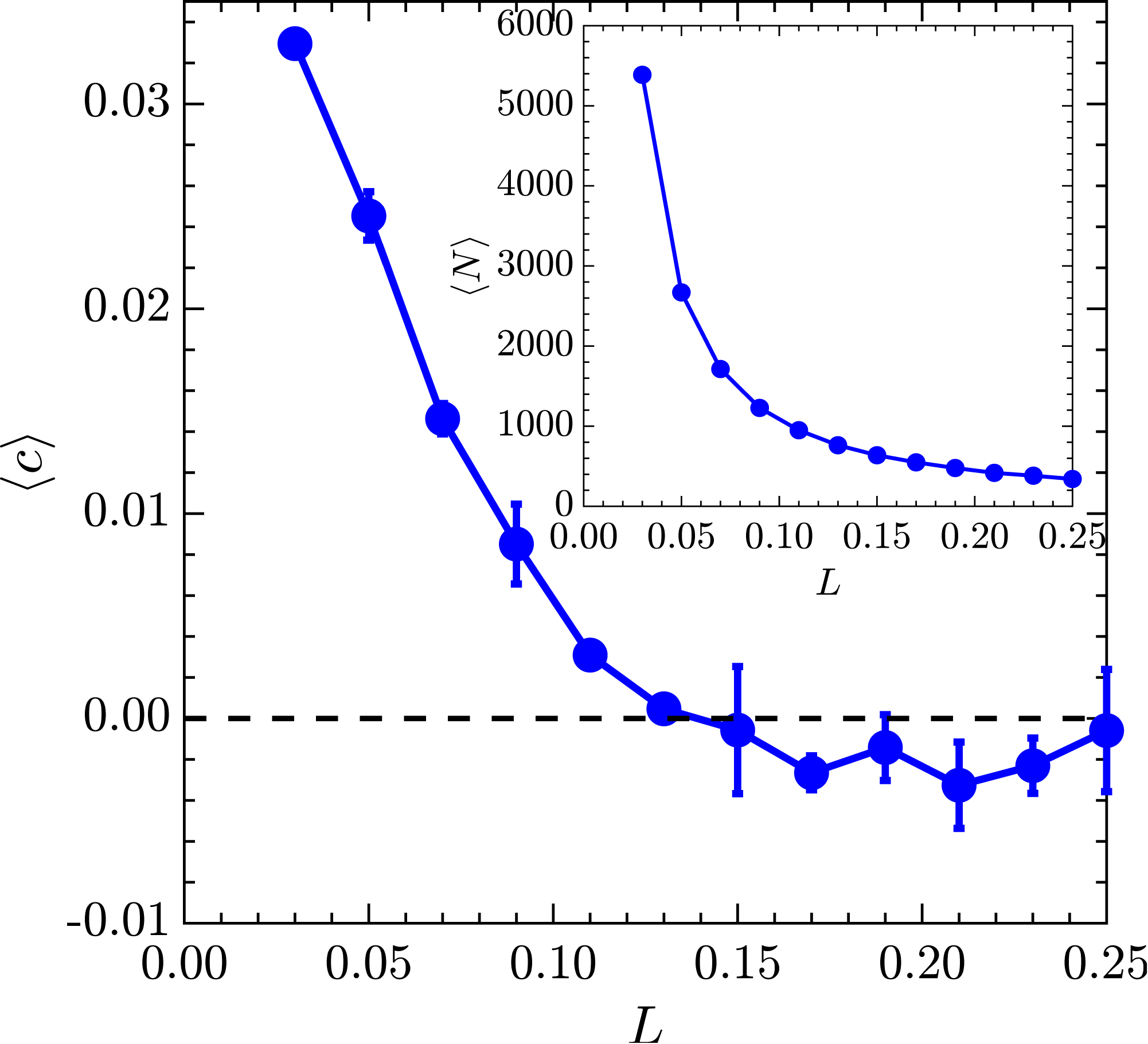}
    \caption{\textbf{Effect of orientational elasticity, $L$:} Variation in the charge of the active fluid to changes in the orientational elasticity for $\zeta = 0.15$ and $K = 0.1$. Inset shows the variation in the number of defects with change in the orientational elasticity.}
    \label{fig:L_effect}
\end{figure}

The resistance to the deformation of the nematic structure observed in bacterial colonies and epithelial cell layers depends on the orientational elasticity of the material, characterized by the orientational elasticity constant $L$ in the model \cite{bagnani2021elastic}. Figure~\ref{fig:L_effect} describes that the charge of the active fluid decreases with an increase in the orientational elasticity. This is attributed to the increasing energy cost of topological distortions as the orientational elasticity increases, leading to a reduction in the total number of defects within the system, as depicted in the inset of Fig.~\ref{fig:L_effect}. Consequently, it can be inferred that the activity of the fluid effectively decreases with an increase in the orientational elasticity, resulting in a decrease in the charge of the active fluid.

 \subsubsection{Effect of isotropic friction}
 
\begin{figure} [!htb]
    \centering
    \includegraphics[width=0.80\linewidth]{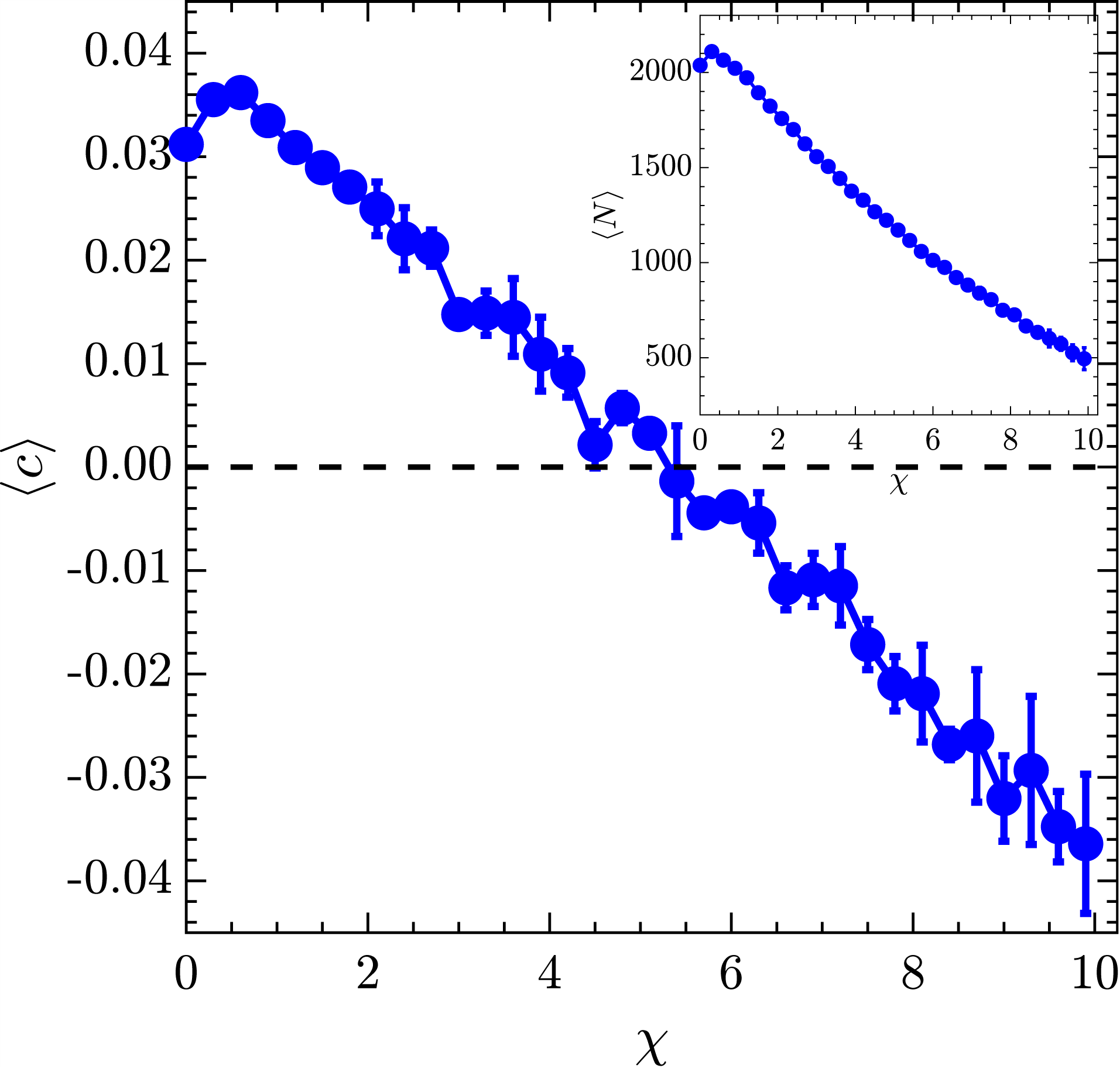}
    \caption{\textbf{Effect of isotropic friction coefficient, $\chi$:} The charge of the active fluid changes with increase in the isotropic friction coefficient for $\zeta = 0.15$ and $K = 0.05$. Inset shows the variation of the number of defects with the strength of the isotropic friction.}
    \label{fig:chi_effect}
\end{figure} 

Figure \ref{fig:chi_effect} plots the charge of the active fluid as a function of the friction coefficient. The charge exhibits a non-monotonic variation with the friction coefficient, initially increasing before subsequently decreasing.  This behaviour can be attributed to the corresponding variation in the total number of defects within the system, as depicted in the inset of Fig. \ref{fig:chi_effect}. Such behaviour mirrors that observed in a single nematic system~\cite{thampi2014active}, where the number of defects initially increases with friction due to an increase in the number of walls decaying into defects. However, at higher levels of frictional damping, there is not enough energy available to create defects, resulting in a subsequent decrease in their number. Furthermore, the increase in friction reduces the velocity-velocity correlations~\cite{doostmohammadi2016stabilization,rozman2023dry}, affecting the efficacy of streak propagation and the merging process involving the $+1/2$ defects (Fig.~\ref{fig:pd_mech} (e)--(f)). Additionally, strong friction reduces the mobility of $+1/2$ defects in the passive fluid more than in the active fluid, causing the $+1/2$ defects to localize in the passive fluid, thus resulting in a negatively charged active fluid. In summary, akin to the effect of anchoring strength and orientational elasticity, an increase in isotropic friction decreases the charge of the active fluid. This suggests that the impact of the model parameters on charge segregation aligns with their influence on the activity and number of topological defects.

\subsection{Effect of the active fluid's volume fraction on its charge}

\label{sec:conc_effect}

 \begin{figure*} [htb!]
	\centering	\includegraphics[width=0.95\linewidth]{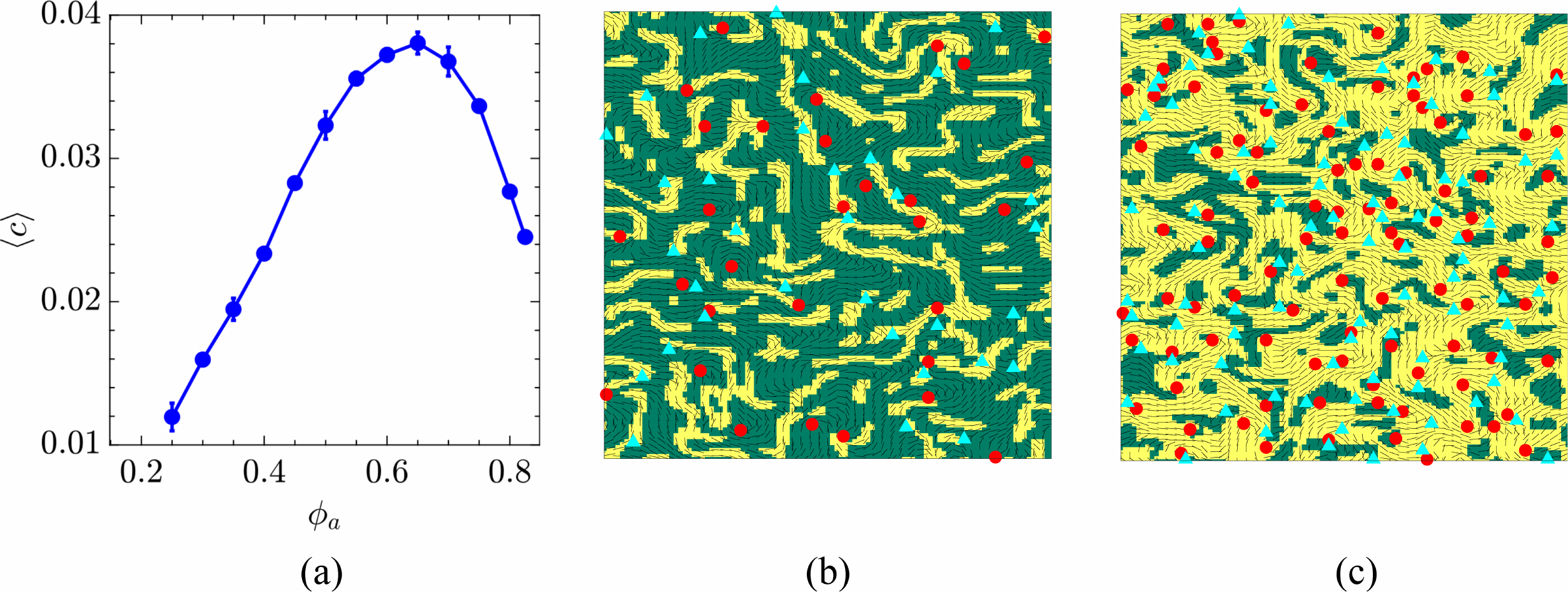}
	\caption{\textbf{Effect of active fluid's volume fraction:} (a) The variation in the charge of the active fluid with its volume fraction for $\zeta = 0.15$ and $K = 0.05$. The instantaneous snapshot of the phase field variable ($\phi$) with overlaid director field (black lines) and topological defects ($
 1/2$ depicted as red circles, $-1/2$ defects as cyan triangles) for (b) $\phi_a = 0.3$ and (c) $\phi_a = 0.7$.}
	\label{fig:Concentration_effect}
\end{figure*}

Finally, we explore how the charge segregation is affected by different proportions of passive to active fluids. The volume fraction of the active fluid ($\phi_a$) is determined by the initial value of the order parameter via $\phi_a = \phi(t=0)$. As shown in Figure~\ref{fig:Concentration_effect}, the charge of the active fluid exhibits a non-monotonic variation with the volume fraction of the active fluid. Specifically, the charge of the active fluid increases with an increase in $\phi_a$ until $\phi_a \approx 0.6$, after which it decreases. Thus, the maximum charge of the active fluid is observed when there is a minimal difference in the volume fraction of active and passive fluids.

The charge of the active fluid is positive for $\phi_a = 0.5$, indicating that the fraction of $+1/2$ defects in the active fluid is more than that of $-1/2$ defects. When $\phi_a<0.5$, the decrease in the amount of active fluid limits the formation of a continuous network, as depicted in Fig.~\ref{fig:Concentration_effect}(b). Consequently, more $+1/2$ defects move into the passive fluid compared to the case of $\phi_a = 0.5$, leading to a decrease in charge. Conversely, when $\phi_a > 0.5$, the higher amount of active fluid facilitates the easier formation of a continuous network, resulting in an increase in the charge of the active fluid. However, when $\phi_a \to 0$ 
, the total number of defects decreases due to a decrease in the fraction of active fluid, making the charge zero, and when $\phi_a \to 1$, both $+1/2$ and $-1/2$ defects predominantly remain within the active fluid as shown in Fig.~\ref{fig:Concentration_effect}(c), suppressing the charge segregation and decreasing the active fluid's charge.  Thus, the mechanisms driving the decrease in active fluid charge for low- and high-volume fractions of the active fluid are distinct.

\section{Conclusions and outlook}

In this study, we investigated the dynamics of charge segregation and defect transport in a binary mixture of active-passive nematic fluids using a biphasic nematic framework. Our findings show that the phase separation of active and passive fluids is a dynamic process characterized by the continuous breaking and reforming of phase-separated domains. 

While, half-integer defects nucleate and annihilate in pairs, maintaining charge neutrality within the binary mixture, our results show that their uneven distribution between active and passive fluids can disrupt the charge balance within each fluid, leading to charge segregation. Our findings indicate that increasing activity leads to an increase in the charge of the active fluid, reflecting a predominance of $+1/2$ defects in the active fluid. Conversely, increasing the interfacial tension decreases the charge of the active fluid, indicating the exit of $+1/2$ defects from the active fluid. Therefore, the active fluid becomes positively charged when activity is high and interfacial tension is low. This happens because the $+1/2$ defects primarily move within the active fluid due to the formation of a constantly reforming active fluid network, whereas the flow around $-1/2$ defects drive away the active fluid. In contrast, when interfacial tension dominates over activity, $+1/2$ defects migrate from the active fluid to the passive fluid, while $-1/2$ defects tend to remain within its bulk, resulting in a negatively charged active fluid.

Furthermore, we examined the impact of interfacial anchoring on charge segregation. The charge of the active fluid is found to decrease with increasing anchoring strength. This is attributed to the enhanced alignment of the director field with the interface, resulting in higher nematic order and fewer defects. This results in a decrease in the charge of the active fluid. Additionally, through analysis of the orientational elasticity and isotropic friction, we demonstrated that the impact of these model parameters on charge segregation aligns with their influence on activity and the number of topological defects.  Moreover, we observed that the charge of the active fluid exhibits a non-monotonic variation with the volume fraction of active fluid, with distinct mechanisms driving charge segregation for low and high-volume fractions.

This study highlights the combined influence of activity and interfacial tension on charge segregation in a binary mixture of active and passive fluids. Furthermore, by analysing the effect of interfacial anchoring, orientational elasticity, isotropic friction, and volume fraction, our findings offer new insights into the diverse factors that control the behaviour of these mixed nematic systems.  Motile and non-motile cells routinely encounter one another in both the natural environment and the human body \cite{yang2004high, vos2008natural, clark2015phenotypic,kwon2019stochastic,zanotelli2022highly}. Our simulations therefore lay the foundation for understanding the emergent physical processes that structure these microscale ecosystems, which ultimately might lead to new ways to control them.

\newpage

\section*{Author Contribution}
C. KVS., A. A., S. T., and A. D. designed and performed the research. C. KVS wrote the first draft of the manuscript. All authors analyzed the results and contributed to editing the manuscript.

\section*{Conflicts of interest}
There are no conflicts to declare.

\section*{Acknowledgements}
C. KVS acknowledges the Post-Doctoral Equivalent Fellowship from the Indian Institute of Technology Madras for supporting this research. A. A. acknowledges support from the EU’s Horizon Europe research and innovation program under the Marie Sklodowska-Curie grant agreement No. 101063870 (TopCellComm). A. D. acknowledges funding from the Novo Nordisk Foundation (grant No. NNF18SA0035142 and NERD grant No. NNF21OC0068687), Villum Fonden (Grant no. 29476), and the European Union (ERC, PhysCoMeT, 101041418).
 Views and opinions expressed are however those of the authors only and do not necessarily reflect those of the European Union or the European Research Council. Neither the European Union nor the granting authority can be held responsible for them.





\bibliographystyle{apsrev4-1}       

\providecommand{\noopsort}[1]{}\providecommand{\singleletter}[1]{#1}%
\providecommand*{\mcitethebibliography}{\thebibliography}
\csname @ifundefined\endcsname{endmcitethebibliography}
{\let\endmcitethebibliography\endthebibliography}{}

\end{document}